\documentclass[aps,prl,twocolumn,superscriptaddress,showpacs,showkeys,amsfonts,amssymb,amsmath]{revtex4-1}\usepackage{graphicx}
\usepackage{ulem}
\usepackage[colorlinks=true,linkcolor=blue,citecolor=blue,urlcolor=blue]{hyperref}
\usepackage{float}
\usepackage{xspace}
\usepackage[T1]{fontenc}
\usepackage{array}
\usepackage{tabularx}
\newcommand{\degree}{\ensuremath{^\circ}\xspace}
\newcommand{\TTO}{Tb$_2$Ti$_2$O$_7$\xspace}
\newcommand{\wvnmbr}{cm$^{-1}$\xspace}

\newcommand{\IIO}{\ensuremath{[110]}\xspace}

\newcommand{\hoti}{Ho$_2$Ti$_2$O$_7$}
\newcommand{\tbti}{Tb$_2$Ti$_2$O$_7$}

\newcommand{\dyti}{Dy$_2$Ti$_2$O$_7$}

\newcommand{\ho}{Ho$^{3+}$}
\newcommand{\tb}{Tb$^{3+}$}
\newcommand{\dy}{Dy$^{3+}$}

\begin{document}

\title{Double vibronic process in the quantum spin ice candidate Tb$_2$Ti$_2$O$_7$ revealed by terahertz spectroscopy}

\author{E. Constable}
\email[]{evan.constable@neel.cnrs.fr}
\author {R. Ballou}
\author{J. Robert}
\affiliation{Institut N\'eel, CNRS and Universit\'e Grenoble Alpes, F-38042 Grenoble, France}
\author{C. Decorse}
\affiliation{ ICMMO, Universit\'e Paris-Sud, F-91400 Orsay, France}
\author{J.-B. Brubach}
\author{P. Roy}
\affiliation{Synchrotron SOLEIL, L'Orme des Merisiers, Saint-Aubin, BP 48, F-91192 Gif-sur-Yvette Cedex, France}
\author {E. Lhotel}
\affiliation{Institut N\'eel, CNRS and Universit\'e Grenoble Alpes, F-38042 Grenoble, France}
\author {L. Del-Rey}
\affiliation{Institut N\'eel, CNRS and Universit\'e Grenoble Alpes, F-38042 Grenoble, France}
\author {V. Simonet}
\affiliation{Institut N\'eel, CNRS and Universit\'e Grenoble Alpes, F-38042 Grenoble, France}
\author{S. Petit}
\affiliation{Laboratoire L\'eon Brillouin, CEA, CNRS, Universit\'e Paris-Saclay, CEA Saclay, F-91191 Gif-sur-Yvette Cedex, France}
\author{S. deBrion}\email[]{sophie.debrion@neel.cnrs.fr}
\affiliation{Institut N\'eel, CNRS and Universit\'e Grenoble Alpes, F-38042 Grenoble, France}

\date{\today}

\begin{abstract}
The origin of quantum fluctuations responsible for the spin liquid state in \tbti\, has remained a long standing problem. By synchrotron-based terahertz measurements, we show evidence of strong coupling between the magnetic and lattice degrees of freedom that provides a path to the quantum melting process.
As revealed by hybrid crystal electric field-phonon excitations that appear at 0.67\,THz below 200\,K, and around 0.42\,THz below 50\,K, the double vibronic process is unique for Tb$^{3+}$ in the titanate family due to adequate energy matching and strong quadrupolar moments. The results suggest that lattice motion can indeed be the driving force behind spin flips within the hybridized ground and first excited states, promoting quantum terms in the effective Hamiltonian that describes \tbti.
\end{abstract}.



\maketitle

\begin{figure}[h]
\resizebox{8.6cm}{!}{\includegraphics{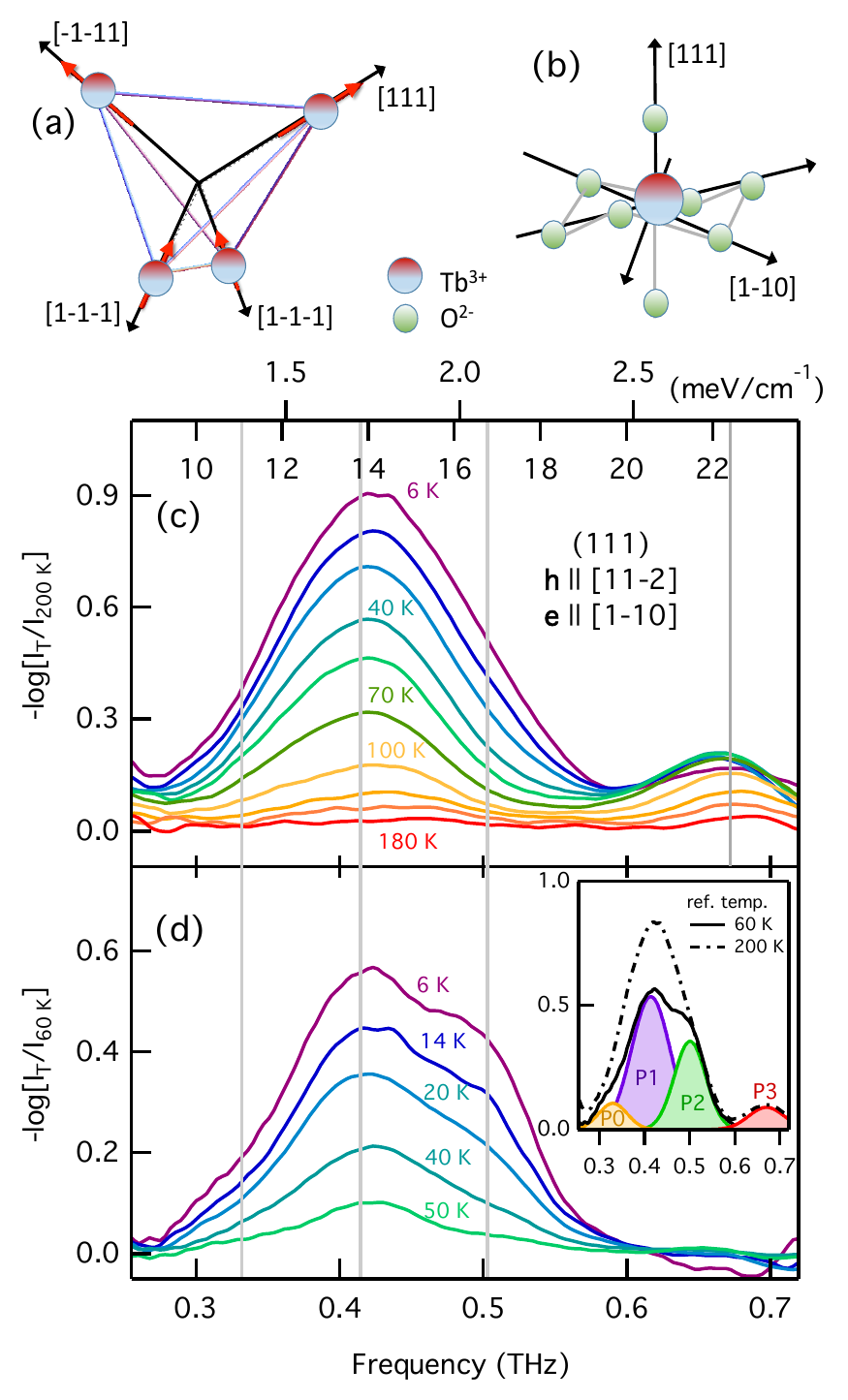}}
\caption{Local symmetries and THz spectra close to the first CEF excitation in Tb$_2$Ti$_2$O$_7$. (a) One tetrahedron showing the ``two-in two-out" spin configuration with the four spin directions. (b) Local $D_{3d}$ environment of each magnetic ion. (c, d) Temperature dependent THz spectra highlighting the two temperature regimes by using the sample as a reference at 200\,K (c) and 60\,K (d). The spectra are fitted by four Gaussian profiles designated P0, P1, P2 and P3 (inset). Vertical gray lines indicate base temperature positions of fitted peaks.}
\label{fig:AbsSpec}
\end{figure}

Over the last decade, spin-ice physics has aroused significant attention \cite{Lhuillier11,Gingras10}.
It emerges in pyrochlore magnets, such as \hoti\, or \dyti, as a result of effective ferromagnetic interactions coupling the Ising-like rare-earth magnetic moments constrained along the local $\langle111\rangle$ directions by the crystal field \cite{Harris97,Ramirez99,Denhertog00} (Fig. \ref{fig:AbsSpec} (a,b)). This combination impedes long-range order, favoring a disordered and macroscopically degenerate magnetic state governed by an organizing principle, the so-called ``ice rule," where two spins point into and two out of each elementary tetrahedron in the structure \cite{Harris97}.

The possibility that quantum fluctuations can melt the spin ice state is a topical issue. It is expected to lead to a wealth of exotic phenomena such as a quantum superposition of ``two-in two-out'' configurations and emergent electrodynamics with new deconfined particles \cite{Balents04,Onoda11,Savary12,Shannon12,Sungbin12,Gingras14}.

In this context, the puzzling behavior of the spin liquid \tbti\ has attracted much attention. Despite effective antiferromagnetic interactions, \tbti\ features no long-range order \cite{Gardner99,Gingras00}, however short-range correlations are present over a broad temperature range \cite{Fritsch14,Kermarrec15,Guitteny15}. Remarkably, the ground state supports elastic power law spin correlations \cite{Fennell12,Petit12}, bearing some resemblance to the pinch point pattern observed in spin ices \cite{Fennell09}. Below 300\,mK, a dispersing low-energy collective mode is found to emerge from these pinch points, reaching its maximum energy at $\sim$0.25\,meV \cite{Guitteny13}. This signal likely corresponds to fluctuations between the two $|\pm\rangle$ electronic states of the crystal electric field (CEF) ground state doublet. It was interpreted as a signature of interactions between quadrupoles, highlighting the importance of those degrees of freedom \cite{Petit12,Curnoe2008,Guitteny13,Takatsu-prl-16}. This is because the non-Kramers nature of Tb$^{3+}$ ions prevents magnetic exchange alone from inducing such fluctuations within the Ising-like ground state \cite{Mueller1968,Curnoe2008,Bonville11}.

A property that distinguishes \TTO from the other rare-earth titanates is the CEF energy spectrum of Tb$^{3+}$. It features a first excited doublet at a low energy of $\Delta \approx 1.5$\,meV above the ground state \cite{Cao09,Zhang14,Princep15,Ruminy2016_2,Lummen08}, compared to 25\,meV for Ho$^{3+}$ in the prototype spin ice compound.
Furthermore, below $\sim$20\,K, this excited doublet has been shown to couple to a transverse acoustic phonon, forming a hybrid magneto-elastic excitation \cite{Fennell14}. It was proposed that a magneto-elastic coupling, linear in the phonon displacement and in the quadrupolar electronic moments, could explain the peculiar ground state of \tbti. To confirm this hypothesis, it is important to show that this magneto-elastic coupling not only occurs for the excited CEF doublet, but also with the ground doublet.

In this article, we provide evidence for such a coupling using terahertz (THz) spectroscopy, which probes the Brillouin zone center in the energy range close to $\Delta$. Our synchrotron-based experiments, performed with an unprecedented accuracy, show a splitting of both the ground and first excited doublets that occurs by coupling with the acoustic phonon at temperatures as high as 50\,K. This confirms that the actual ground state of \tbti\ has a mixed magnetic and phononic nature. Moreover, we show that this low temperature vibronic process is not alone. A second process is observed as high as 200\,K and couples the first excited CEF state and associated transition at $\Delta$ with a silent optical phonon mode. This double vibronic process provides a natural path for spin-lattice quadrupolar coupling that amplifies transverse contributions on top of the Ising anisotropy.

The measurements were performed using a \tbti\, single crystal grown by the floating zone technique previously described in \cite{Petit12,Guitteny13}. We have measured the polarization dependent THz spectrum in the 0.2--1.0\,THz (0.82--4.14\,meV) energy range using both standard (SSR) \cite{Chaix13} and coherent synchrotron radiation (CSR) \cite{Barros13, Barros15} techniques. Focus is given to the ``high temperature'' regime, from 300\,K to 6\,K. Four plaquettes, 2\,mm in diameter, were chosen to test the magnetoelectric activity of the excited modes according to the main cubic symmetry axes. Specifically, the (100) wafer  (1\,mm thick) is perpendicular to the four-fold axis while the (111) wafer (300\,$\mu$m thick) is perpendicular to the three-fold axis. To probe the local rare earth symmetry with an Ising direction along $[111]$, two additional wafers (1$\bar{1}$0), and (11$\bar{2}$) (300\,$\mu$m thick) are used allowing three mutually orthogonal planes: (111), (1$\bar{1}$0), and (11$\bar{2}$).
Note then, 
that the projection of the polarized THz field in the local frame is mixed between the 4 Tb$^{3+}$ sites of each tetrahedron (See Supplemental Material \cite{supmat}).

\begin{figure}
\resizebox{8.6 cm}{!}{\includegraphics{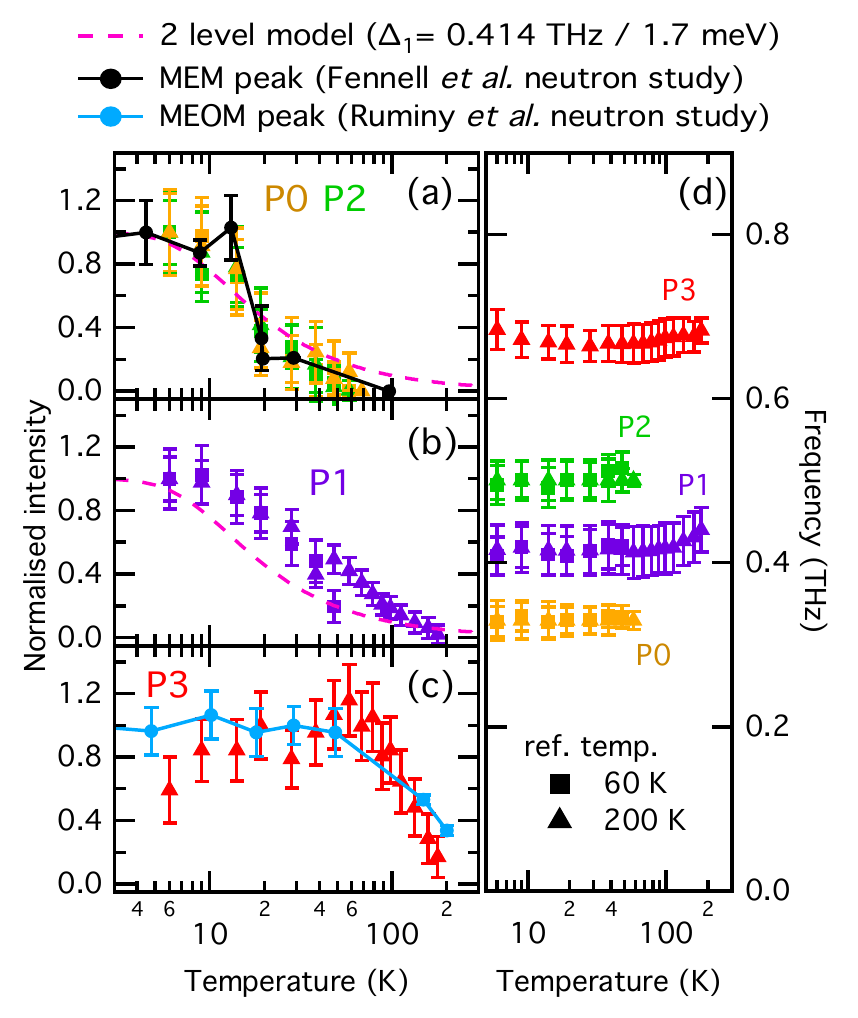}}
\caption{(a-c) Temperature dependence of the P0, P1, P2 and P3 peak intensity determined by the area of the Gaussian fits. Peak area is normalized to the value at 6\,K. In panel (a) the intensity of the MEM measured by Fennell \textit{et al.} \cite{Fennell14} at 5 meV is also shown, which represents the onset of magneto-elastic correlations. The population profile of a two-level model, $n_1-n_0$ where $n_i=\exp(-\Delta_i/k_BT)/\sum_0^1\exp(-\Delta_j/k_BT)$ and $\Delta_1=1.7$\,meV, is added in (a) and (b). In panel (c) the intensity of the MEOM measured by Ruminy \textit{et al.} \cite{Ruminy2016_2} at 16 meV represents the onset of high temperature CEF-Phonon coupling. (d) Temperature dependence of fitted peak energies. Uncertainties represent the span of the peak's half width at half maximum.}
\label{fig:TmpDpnd}
\end{figure}

Typical spectra are given in Fig. \ref{fig:AbsSpec}(c) with a 200\,K reference. As the temperature is lowered down to 6\,K, an absorption peak develops around 0.42\,THz (1.7\,meV). This feature is consistent with the first excited CEF level previously reported by neutron and optical spectroscopic studies \cite{Cao09,Zhang14,Princep15,Ruminy2016_2,Lummen08}. A much weaker and so far unreported peak is visible at 0.67\,THz (2.78\,meV). Previous experimental studies suggest that the low-temperature magnetic correlations first begin to develop below $\sim$100\,K with short-range correlations strengthening below 20\,K \cite{Gardner99,Gardner01}. We therefore analyze this temperature regime using the spectrum at 60\,K (Fig. \ref{fig:AbsSpec}(d)) as a reference. Noticeably, in Fig. \ref{fig:AbsSpec}(d) the peak at 0.67\,THz (2.78\,meV) is not present indicating that its strength and energy position do not change significantly over this temperature range and its contribution is divided out along with the background. At the same time, close inspection reveals a three peak structure within the CEF excitation. The four observed peaks, with 200\,K and 60\,K references, are fit by Gaussian profiles shown in the inset of Fig.~\ref{fig:AbsSpec}(d) designated P0, P1, P2 and P3. Their energy position and normalized total area are shown in Fig. \ref{fig:TmpDpnd}. No significant temperature shifts are observed for their energy position except for a slight hardening below 20\,K for P3 and above 100\,K for P1. The precise fitted peak energies at 6\,K are E0\,=\,0.332(4)\,THz (1.37(2)\,meV), E1\,=\,0.415(6)\,THz (1.71(2)\,meV), E2\,=\,0.502(9)\,THz (2.08(3)\,meV), and E3\,=\,0.672(6)\,THz (2.78(3)\,meV).
P0, P1 and P2 mainly appear below 50\,K, while P3 appears at the much higher temperature of 200\,K, indicating a different origin.

\begin{figure}[h]
\resizebox{8.6 cm}{!}{\includegraphics{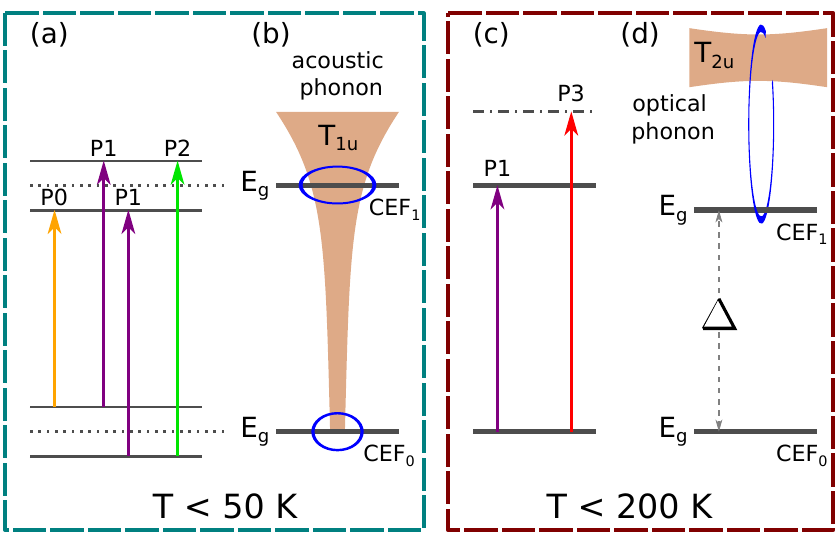}}
\caption{ Sketch of the two CEF-phonon hybridization processes that occur successively. (a) Four level scheme giving rise to the three excitations P0-P2. (b) Low temperature coupling process involving two CEF states and an acoustic $T_{1u}$ phonon. (c) Two level scheme giving rise to excitations P1 and P3. (d) High temperature coupling process involving the first excited CEF state and a low-energy $T_{2u}$ optical phonon.}
\label{fig:Hyb}
\end{figure}

It is tempting to assign the three peak structure (P0-P2) to a splitting of the ground and first excited CEF levels, that share the $E_g$ symmetry. The resulting four level scheme will give rise to the 3 observed excitations if the central peak (P1) corresponds to two different transitions of the same energy (see Fig.~\ref{fig:Hyb} (a)) within our experimental precision. The $E_g$ splitting is then similar for both the ground state doublet and the first excited doublet at a value of 0.085\,THz (0.35\,meV). Such a splitting corresponds very nicely in energy with the collective mode measured by neutron scattering at a much lower temperature (50\,mK) \cite{Guitteny13}.
It should be emphasized that it is experimentally impossible to assign the entire four peak spectrum to transitions between the two doublets with different splittings \cite{supmat}.

The temperature profile for the outer peaks (P0 and P2) is modeled by a two-level system with an energy barrier of 0.42\,THz ($\Delta\approx$ 1.7\,meV). It bears similarity to the temperature dependence of the magneto-elastic mode (MEM) measured by Fennell \textit{et al.} using inelastic neutron scattering \cite{Fennell14,Guitteny13}. This MEM has been shown to arise from hybridization between the $T_{1u}$ acoustic phonon and first excited CEF level taking on a magneto-elastic component. A zone center extrapolation of the scattering vector dependent neutron data is in good agreement with the energy of P2 at 0.50\,THz (2.08\,meV). Since the acoustic phonon branch must also cross the CEF ground state at the zone center, hybridization with the ground state is also expected due to its identical $E_g$ symmetry. This is exactly what we observe, with similar splitting for both the ground and first excited states. This suggests that the collective mode built on the basis of the $|\pm\rangle$ states in the ground doublet already incorporates lattice degrees of freedom as does the first excited CEF state.
Similar vibronic phonon-CEF interactions have also been shown for CeAl$_2$ and facilitate mixing between the ground and first excited CEF states \cite{Thalmeier84}.

In \TTO, the hybridization process between the two $E_g$ CEF levels and the $T_{1u}$ acoustic phonon depicted in Fig.~\ref{fig:Hyb} (b) is not alone. A recent report by Ruminy \textit{et al.} \cite{Ruminy2016_2} suggests that other hybridized magneto-elastic optical mode (MEOM) excitations occur at much higher energies around 3.9\,THz (16\,meV) due to the proximity of optical phonons. Their work is corroborated by a complementary study focused on the phonon spectrum in these compounds \cite{Ruminy2016_3}. Noticeably, their \textit{ab inito} calculations show that the lowest-energy phonon is at $\sim$1.1\,THz (4.6\,mev) and has $T_{2u}$ symmetry. This phonon is optically silent (both in Raman and infrared (IR) spectroscopy), and it is not observed by inelastic neutron scattering because of a negligible scattering cross section. Since it is quite close in energy to the additional P3 excitation that occurs below 200\,K at 0.67\,THz (2.78\,meV), we assign P3 to a coupling mechanism between this $T_{2u}$ phonon and the first excited CEF level. It is worth noting that this optical phonon branch does not cross the $E_g$ ground state, and therefore is not coupled to it.

Using infrared reflection measurements, we have checked that the 7 expected IR active phonon modes are present in the range 80--800\, \wvnmbr (2.4--24\,THz) \cite{supmat}.
The observed phonon energies reveal a discrepancy in comparison to the calculated values which suggests that the lowest-energy $T_{2u}$ phonon may indeed be even closer to the first excited $E_g$ CEF level.

Therefore, two different vibronic processes must occur in \TTO at two different temperature scales. A diagram of the two coupling mechanisms and corresponding level scheme is depicted in Fig.~\ref{fig:Hyb}. The high-temperature vibronic process involves the first excited $E_g$ CEF and the lowest energy $T_{2u}$ optical phonon. The resulting hybridization develops below 200\,K as seen with the appearance of peak P3 at an intermediate energy between the $E_g$ CEF level and the $T_{2u}$ phonon energy. This coupling mechanism transfers magnetic THz activity to an otherwise silent phonon. The low temperature vibronic process occurs when magnetic fluctuations set in below 50\,K. It involves the ground and first excited CEF levels, which both have $E_g$ symmetry, and the $T_{1u}$ acoustic phonon. It results in the splitting of both the ground and first excited doublets which acquire a mixed electronic-phononic character.

\begin{figure} 
\resizebox{8.6 cm}{!}{\includegraphics{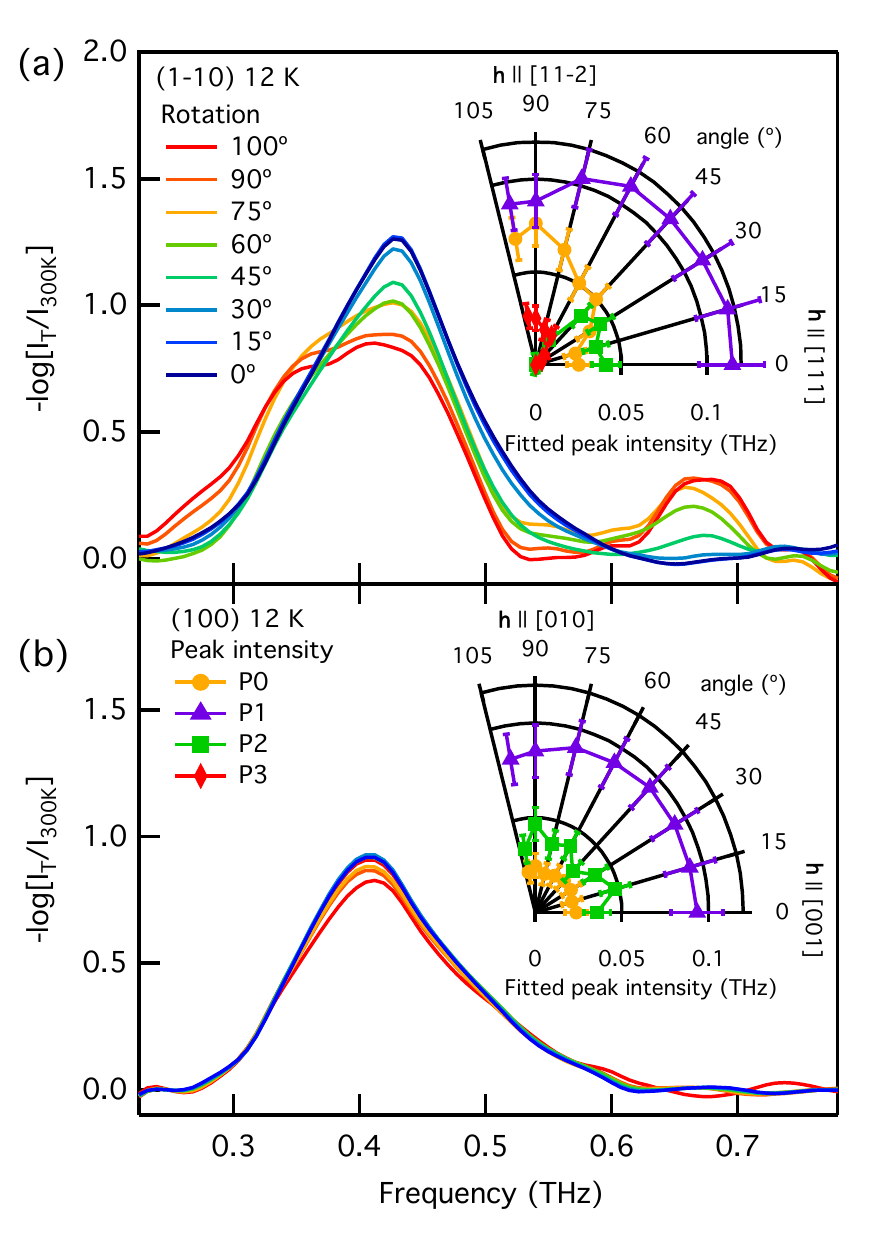}}
\caption{Angular dependence of \TTO spectra around the \IIO (a) and [100] (b) rotation axes obtained using CSR configuration. Insets show fitted peak intensities as a function of the THz magnetic field orientation.}
\label{fig:AnglDpnd}
\end{figure}

Additional valuable information about these CEF-phonon processes is obtained by the selection rules observed for each excitation and their angular dependence when the polarization of the THz electric ($\mathbf{e}$) and magnetic ($\mathbf{h}$) fields is rotated. We first confirm that the hybridized features (P2 and P3) are mainly of magnetic origin. When measuring two configurations with the same direction for $\mathbf{h}$ but 90 degrees rotation for $\mathbf{e}$, the same features are present \cite{supmat}.
Using a rotative sample mount, we were then able to precisely measure the angular dependence of the spectrum for a rotation about the [100] fourfold and $[1\bar{1}0]$ two fold axes of the cubic structure (Fig. \ref{fig:AnglDpnd}). Note that the [100] axis has no particular meaning in the rare earth site symmetry while the $[1\bar{1}0]$ direction corresponds to a two fold axis for two sites out of four on each tetrahedron \cite{supmat}.
Around [100], the spectrum remains unchanged as expected. In contrast, around $[1\bar{1}0]$, a strong anisotropy appears for all peaks except P1. P2 has a maximum intensity for $\mathbf{h} \parallel [111]$, while both P0 and P3 behave inversely with maximum intensities when $\mathbf{h} \bot [111]$. 
There is also a transfer of spectral weight between P2 and P3 showing that the two vibronic processes are not independent.

This double vibronic mechanism accounts for a mixing between electronic states and propagating phonon modes, which occur in crystals when the excitations are of the same symmetry and are close enough in energy \cite{supmat}. The transverse components of the acoustic T$_{1u}$ phonon fulfill the symmetry requirements for both the ground and first excited CEF states and their coupling may induce a splitting of these doublet levels. As for the silent optical T$_{2u}$ phonon, its longitudinal component is also allowed to  couple to the CEF states inducing no splitting but a shift in the CEF level. This  optical phonon is close in energy to the first excited CEF doublet and will dominantly form vibrons with it at high temperature. Its dispersion can be reasonably assumed to be flatter than that of the acoustic phonon leading to a larger density of states. At lower temperatures, this vibronic process should weaken owing to thermal depopulation. The remaining acoustic-phonon vibronic process will then gradually take hold.
While P3 is still present at 6\,K, this process appears evident in Fig.~\ref{fig:TmpDpnd} (c) with a slight reduction in the intensity of P3 at the lowest temperatures.

Importantly, the double vibronic process involves quadrupolar operators which behave as 
 spin flip operators in the $|\pm \rangle$ ground state doublet basis. As a result, phonons (and lattice zero point motions) are now the source of spin fluctuations and by their virtual exchange may give rise to new quantum terms in the effective Hamiltonian of \tbti. 
 Depending on their strength, a long-range quadrupolar order or Coulomb-phase--like state might be stabilized, provided the Ising term 
 dominates \cite{Takatsu-prl-16,Onoda11,Sungbin12}. Furthermore, an interesting feature of the vibronic process is that the acoustic phonon crosses the two doublets, which themselves being close in energy, will also partially mix. We anticipate that the magnetic dipolar operators will then behave as spin flip operators 
 in the basis of the dressed $|\pm\rangle$ electronic states. This mechanism, already proposed in Ref \cite{Molavian07}, could be the source of additional fluctuations. It may also be present in Tb$_2$Sn$_2$O$_7$ but is not allowed in \hoti\, and \dyti: Although the symmetries of both phonons and CEF states are identical, the energy of the first CEF state is one order of magnitude larger, reducing the vibronic process solely to the ground state. More importantly, the quadrupolar matrix elements in the basis of the ground $|\pm\rangle$ electronic states are essentially zero for \ho\, and \dy.

In summary, using synchrotron-based THz spectroscopy, we show that two temperature regimes are present in the spin ice candidate \TTO\ revealing two different vibronic processes. The first one, below 50\,K, concerns the ground and first excited CEF levels which couple to the acoustic $T_{1u}$ phonon. This is the temperature scale of magnetic fluctuations \cite{Gardner99} as well as elastic softening \cite{Nakanishi11}. The second regime, below 200\,K, is associated with the coupling between the first CEF level and the lowest energy optical phonon of symmetry $T_{2u}$ and is in agreement with the high temperature plateau observed in the muon spin relaxation rate \cite{Gardner99} for this compound.
This double vibronic process, of quadrupolar origin, affects the electronic state that can no longer be described solely by electronic wave functions. Rather, a collective vibronic state prevails, built on the ground and first excited CEF states mixed with two different phonons. It is activated with dramatic effects in \TTO\ by virtue of adequate energy matching and strong quadrupolar moments that occur in this compound.
This unique combination of vibronic processes revealed by our THz study supports the formation of the \tbti\, quantum spin liquid state; it may also provide a broadly applicable model in the formation of other exotic quantum ground states.

\section {Acknowledgements}
This work was supported by the French ANR Project No. ANR-13-BS04-0013-01. We thank J. Debray for sample preparation and F. L\'{e}vy-Bertrand for assistance with infrared measurements.


%
%
%

\onecolumngrid
\pagebreak
\renewcommand{\thefigure}{S\arabic{figure}}
\renewcommand{\thetable}{T\arabic{table}}
\section{Supplementary Material}
\subsection{Geometry of the rare earth stacking}
\begin{figure}[H]
\centering
\includegraphics[width=0.5\linewidth]{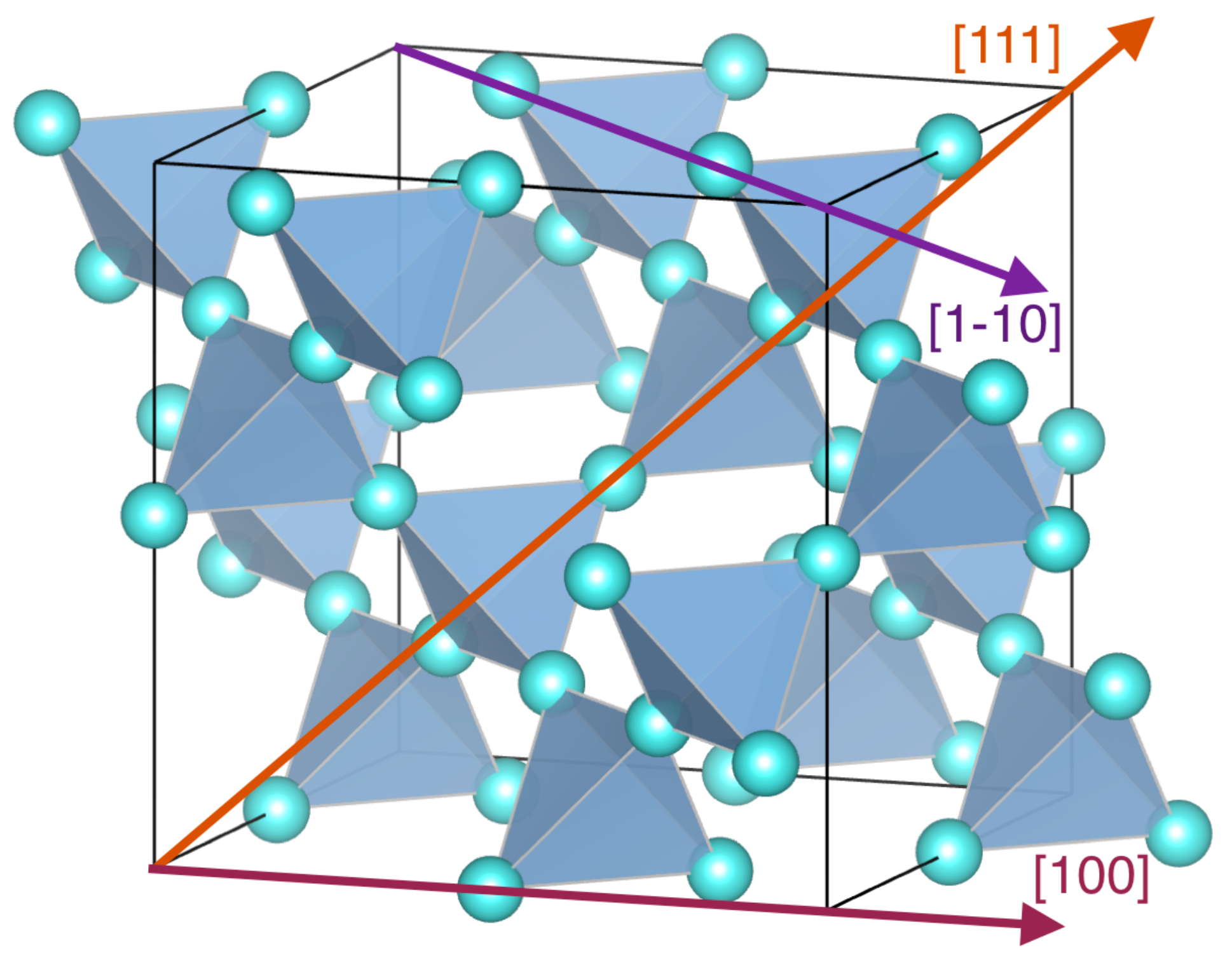}
\caption{The pyrochlore network of Tb$^{3+}$ tetrahedra connected at their apexes. The 4-fold [100], 3-fold [111] and 2-fold [110] axes of the cubic $Fd\bar{3}m$ structure are shown. Image generated using VESTA software package \cite{Vesta}.}
\label{fig:pryoLatt}
\end{figure}

\TTO crystallizes in the pyrhochlore structure with space group $Fd\bar{3}m$. The Tb$^{3+}$ ions are the only ones carrying a magnetic moment and occupy the 16d site with local symmetry described by the point group $\bar{3}m$, symbolised $D_{3d}$ in the notation of Schoenflies. The network of apex-connected tetrahedra they form is generated from the Tb$^{3+}$ ions of a single tetrahedron by applying translations of the face centered cubic lattice (Fig.~\ref{fig:pryoLatt}). In each of the four  lattices the local 3-fold axis coincides with one of the four diagonals of the cubic cell. There are three local 2-fold axes perpendicular to each 3-fold axis that coincide with three face diagonals of the cubic lattice. Each 2-fold axis is perpendicular to two 3-fold axes. Calling the z-axis the appropriate cube diagonal and the x-axis one of the 2-fold axes perpendicular to it, we define a local frame on which the THz electric and magnetic field components can be projected. The interrelation between the four local frames and the global cubic frame are provided in table~\ref{tab:LclBs}.

\begin{table}[H]
\centering
\def\arraystretch{1.8}
\caption{The four local bases associated to the four Tb$^{3+}$ ions on each tetrahedron expressed in the cubic frame. The projections of vectors in the cubic frame onto the local frames are also shown.}
\label{tab:LclBs}
\begin{tabular}{r|cccc}
\hline \hline
Local axes & Tb 1 & Tb 2 & Tb 3 & Tb 4 \\ \hline
$x_i$ & $\frac{1}{\sqrt{2}}[1\bar{1}0]$ & $\frac{1}{\sqrt{2}}[\bar{1}10]$ & $\frac{1}{\sqrt{2}}[110]$ & $\frac{1}{\sqrt{2}}[\bar{1}\bar{1}0]$ \\
$y_i$ &$\frac{1}{\sqrt{6}}[11\bar{2}]$  & $\frac{1}{\sqrt{6}}[\bar{1}\bar{1}\bar{2}]$ & $\frac{1}{\sqrt{6}}[1\bar{1}2]$ & $\frac{1}{\sqrt{6}}[\bar{1}12]$ \\
$z_i$ & $\frac{1}{\sqrt{3}}[111]$ & $\frac{1}{\sqrt{3}}[\bar{1}\bar{1}1]$ & $\frac{1}{\sqrt{3}}[1\bar{1}\bar{1}]$ & $\frac{1}{\sqrt{3}}[\bar{1}1\bar{1}]$ \\ \hline
$\vec{t}\|[11\bar{2}]$ & $[010]$ & $[0, \frac{1}{3},\frac{-4}{3\sqrt{2}} ]$ & $[\frac{1}{\sqrt{3}}, \frac{-2}{3},\frac{2}{3\sqrt{2}}]$ & $[\frac{-1}{\sqrt{3}}, \frac{-2}{3},\frac{2}{3\sqrt{2}} ]$ \\
$\vec{u}\|[1\bar{1}0]$ & [100] & $[\bar{1}00]$ & $[0, \frac{1}{\sqrt{3}},\frac{2}{\sqrt{6}}]$ & $[0, \frac{-1}{\sqrt{3}},\frac{-2}{\sqrt{6}}]$ \\
$\vec{v}\|[111]$ & [001] & $[0, \frac{-4}{3\sqrt{2}},\frac{-1}{3}]$ & $[\frac{2}{\sqrt{6}}, \frac{2}{3\sqrt{2}},\frac{-1}{3}]$ & $[\frac{2}{\sqrt{6}}, \frac{2}{3\sqrt{2}},\frac{1}{3}]$ \\
$\vec{v}\|[001]$ & $[0, \frac{-\sqrt{2}}{\sqrt{3}},\frac{1}{\sqrt{3}}]$ & $[0, \frac{-\sqrt{2}}{\sqrt{3}},\frac{1}{\sqrt{3}}]$ & $[0, \frac{\sqrt{2}}{\sqrt{3}},\frac{-1}{\sqrt{3}}]$ & $[0, \frac{\sqrt{2}}{\sqrt{3}},\frac{-1}{\sqrt{3}}]$
\\ \hline \hline
\end{tabular}
\end{table}

\subsection{Crystal electric field - phonon coupling}

The CEF spectrum of each \tb\, ion is protected from splitting and mixing by the $D_{3d}$ symmetry of its charge environment. 
When the crystal vibrations are taken into account, the positions of the surrounding ions change in time and the cage they form about the \tb\, ion is dynamically deformed. This results in symmetry breaking of their electrostatic interactions. The latter can be expanded over the spherical harmonics ($\mathcal{Y}_n^m$) on the $f$ electrons angular variables $(\theta_i, \phi_i)$ most generally in the form $\sum_i\sum_{nm} \sum_s {A_n^m}_s~r_i^n\mathcal{Y}_n^m(\theta_i,\phi_i)$, where $r_i$ is the $i-$th $f$ electron radial variable and ${A_n^m}_s$ stands for the contribution associated with the surrounding ion $s$. The expansion complies with the $D_{3d}$ symmetry if and only if it involves invariants belonging to the trivial representation $A_{1g}$ of the point group $D_{3d}$, in which case it leads to the CEF Hamiltonian:
\begin{equation}
\label{cef_hamiltonian}
\mathcal{H}_{\rm CEF} =  A_2^0 ~\langle r^2 \rangle ~\mathcal{C}_2^0 + A_4^0 ~\langle r^4 \rangle ~\mathcal{C}_4^0 + A_4^3 ~\langle r^4 \rangle (\mathcal{C}_4^3 - \mathcal{C}_4^{-3}) + A_6^0 ~\langle r^6 \rangle ~\mathcal{C}_6^0 + A_6^3 ~\langle r^6 \rangle (\mathcal{C}_6^3 - \mathcal{C}_6^{-3}) + A_6^6 ~\langle r^6 \rangle (\mathcal{C}_6^6 + \mathcal{C}_6^{-6}),
\end{equation}
where $A_n^m$ accounts for the contribution of all the surrounding charges, $\langle r^n \rangle$ is the expectation value of $r^n$ over the radial part of the f-electron wave function and $\mathcal{C}_n^m$ are the Wybourne operators. The oscillations of the ions modify the coefficients ${A_n^m}_s$. This can be accounted for by the expansion ${A_n^m}_s + \nabla_{\vec{r_s}}~{A_n^m}_s \cdot \delta \vec{r_s} + \cdots$ provided that the displacement $\delta \vec{r_s}$ of the surrounding ion $s$ from its equilibrium positions relatively to the \tb\, ion is small enough. A single ion magnetoelastic interaction Hamiltonian $\mathcal{H}_{ME}$ then emerges to first order. Non linear displacement effects and cooperative multi-ion magnetoelastic interactions occurring at the higher orders will be ignored. $\mathcal{H}_{\rm ME}$ will involve Wybourne operators that are not present in $\mathcal{H}_{\rm CEF}$ whenever $\sum_s \nabla_{\vec{r_s}}~{A_n^m}_s \cdot \delta \vec{r_s}$ is symmetry breaking.
In order to find these operators it proves convenient to describe the linear displacements by normal modes $u_{\Gamma^-}$ that transform according to the parity odd irreducible representations $\Gamma^-$ of the point group $D_{3d}$ and take advantage of the orthogonality properties of the irreducible representations to write the single ion magnetoelastic interaction Hamiltonian in the form:
\begin{equation}
\label{me_hamiltonian}
\mathcal{H}_{\rm ME} = \sum_{n\Gamma\gamma\gamma\prime} \zeta_{n{\Gamma_{\gamma\gamma\prime}}} ~ u_{\Gamma^-_\gamma} ~\langle r^n \rangle ~\sum_{m} f_{\Gamma^+_{\gamma\prime}}(\mathcal{C}_n^m) = \sum_{n\Gamma\gamma\gamma\prime} \zeta_{n{\Gamma_{\gamma\gamma\prime}}} ~ (a{^\dagger}_{\Gamma^-_\gamma} + a _{\Gamma^-_\gamma}) ~\langle r^n \rangle ~\sum_{m} f_{\Gamma^+_{\gamma\prime}}(\mathcal{C}_n^m),
\end{equation}
where $\zeta_{n{\Gamma_{\gamma\gamma\prime}}}$ is the magnetoelastic coefficient associated with the normal mode $u_{\Gamma^-_\gamma}$ and $a{^\dagger}_{\Gamma^-_\gamma}$ ($a_{\Gamma^-_\gamma}$) the creation (annihilation) operator of a phonon in this mode. $\gamma$ distinguishes the different modes within a multidimensional $\Gamma^-$. $\sum_{m}f_{\Gamma^+_{\gamma\prime}}(\mathcal{C}_n^m)$ is a linear combination of the Wybourne operators that are allowed to couple to lattice vibrations. To determine them, we  look at symmetry consideration including spatial parity conservation. $\zeta_{n{\Gamma_{\gamma\gamma\prime}}}$ is a linear combination of the components of $\nabla_{\vec{r_s}}~{A_n^m}_s$. It is therefore parity odd since the quantity $\nabla_{\vec{r_s}}~{A_n^m}_s \cdot \delta \vec{r_s}$ is parity even as a scalar quantity. Accordingly, the invariants $f_{\Gamma^+_{\gamma\prime}}(\mathcal{C}_n^m)$ must belong to the parity even irreducible representations $\Gamma^+$ of the point group $D_{3d}$ whose character is the same as the one of $\Gamma^-$ on the group elements that are not combined with the space inversion and opposite otherwise. With $\zeta_{n{\Gamma^-}} \sim e^2 / r_s^{2n+2}$ the magnetoelastic coupling energy for displacements in the range $0.01-0.1~\text{\AA}$ is numerically estimated in the order of $0.5-5$~meV for $n=2$, in the order of $0.01-0.1$~meV for $n=4$ and in the order of $0.0005-0.005$~meV for $n=6$. It follows that all the Wybourne operators other than the quadrupolar ones ($n=2$) can be ignored in a first approximation. These quadrupolar operators transform according to the irreducible representation $\mathcal{D}_2^+$ of the rotation group $O(3)$ associated with the angular momentum $n=2$. The subduction to the group $D_{3d}$ of $\mathcal{D}_2^+$ is reduced into ${\mathcal{D}_2^+}_{\downarrow D_{3d}} = \Gamma_1^+ \oplus 2~\Gamma_3^+ = A_{1g} \oplus 2~E_g$. An inspection of the invariants reported in table~\ref{D3d_IRREPS} then informs us that the phonon modes belonging to the irreducible representation $E_{u}$ can couple to the CEF states through the quadrupolar Wybourne operators $\mathcal{C}_2^m \pm \mathcal{C}_2^{-m}~(m=1,2)$. Within the ground multiplet $| JM \rangle$ this transposes itself to possible coupling of the $E_u$ phonon modes with CEF states through the quadrupolar Stevens operators:
\begin{eqnarray*}
O_2^2     &= & J_x^2-J_y^2 \\
O_2^{-2} &= & J_xJ_y+J_yJ_x\\
O_2^1     &= & J_zJ_x+J_xJ_z\\
O_2^{-1} & =& J_zJ_y+J_yJ_z
\end{eqnarray*}
The only other possible coupling is with the phonon mode belonging to the irreducible representation $ A_{1u}$ through the quadrupolar Weybourne operator $\mathcal{C}_2^0$ which transposes to the Stevens operator in the ground multiplet $| JM \rangle$  :
\begin{eqnarray*}
O_2^0     &= & 3J_z^2-J(J+1) \\
\end{eqnarray*}

The acoustic phonons at the Brillouin zone centre in \TTO belong to the irreducible representation $T_{1u}$ of the factor group $m\bar{3}m$. Its subduction to the point group $D_{3d}$ is reduced into ${T_{1u}}_{\downarrow D_{3d}} = A_{2u} \oplus E_u$, meaning that the triply degenerate acoustic modes separate themselves into a non degenerate longitudinal mode ($ A_{2u}$) and doubly degenerate transverse modes ($E_u$) with respect to the local 3-fold axis. These transverse acoustic phonon modes are those that might induce the splitting and the mixing of the ground and first excited CEF doublets that are separated only by about $1.5$ meV. In the subspace of these two doublets the single ion magnetoelastic interaction Hamiltonian $\mathcal{H}_{\rm ME}$ takes the form
\begin{equation}
\label{projected_me_hamiltonian}
\mathcal{H}_{\rm ME} = \sum_{\gamma\gamma\prime}\zeta_{2 {E_u}^{\gamma\gamma\prime} } ~ (a^\dagger_{E_u\gamma} + a _{E_u\gamma}) ~\langle r^2 \rangle ~ \sum_{m = 1,2} \sum_{\substack{s = \pm\\ x=0,1}}  \sum_{\substack{s\prime = \pm\\ x\prime=0,1}} | \psi_s^x\rangle \langle \psi_s^x | ~ f_{E_g\gamma\prime}(\mathcal{C}_2^m)~  | \psi_{s\prime}^{x\prime}\rangle \langle \psi_{s\prime}^{x\prime} |,
\end{equation}
where $|\psi_\pm^0 \rangle$ (resp. $|\psi_\pm^1 \rangle$) are two orthogonal states of the ground (resp. excited) doublet. The 3-fold symmetry implies that they must expand on the basis of the multiplet states $|JM \rangle$ in the form $\sum_{n} \alpha_{M-3n} | M-3n \rangle$ whereas the 2-fold symmetry transforms this into $(-1)^{J-M} \sum_n \alpha_{M-3n} (-1)^n | -M+3n \rangle$. The two states are parallel if $M \equiv 0$ (modulo 3), thus forming a singlet, and orthogonal otherwise, forming a doublet. It follows that:

\begin{eqnarray*}
|\psi_+^x \rangle & = & \alpha_{5}^x |+ 5 \rangle + \alpha_{2}^x | + 2 \rangle + \alpha_{-1}^x | - 1 \rangle + \alpha_{-4}^x | - 4 \rangle, \\
|\psi_-^x \rangle & = & \alpha_{5}^x |- 5 \rangle - \alpha_{2}^x | - 2 \rangle + \alpha_{-1}^x | + 1 \rangle - \alpha_{-4}^x | + 4 \rangle,
\end{eqnarray*}
with $\sum_M \alpha_{M}^x \alpha_{M}^{x'} = 0$ to achieve orthogonality of the various states. The oscillator strengths $\langle \psi_s^x | ~ f_{E_g\gamma\prime}(\mathcal{C}_2^m)~  | \psi_{s\prime}^{x\prime}\rangle$ in eq.\ref{projected_me_hamiltonian} are then evaluated to:
\begin{eqnarray*}
\langle \psi_s^x | ~O_2^1~  | \psi_{s\prime}^{x\prime}\rangle &=& (1/i) \langle \psi_s^x | ~O_2^{-1}~  |\psi_{s\prime}^{x\prime}\rangle \\
&=&(1-\delta_{s, s\prime}) \left( -3\sqrt{5/2} (\alpha_2^x \alpha_{-1}^{x\prime} + \alpha_2^{x\prime} \alpha_{-1}^x) + (9/2)\sqrt{11/2} (\alpha_5^x \alpha_{-4}^{x\prime} + \alpha_5^{x\prime} \alpha_{-4}^x) \right) \\
\langle \psi_s^x | ~O_2^2~  | \psi_{s\prime}^{x\prime}\rangle & =&  (-1/i) \langle \psi_s^x | ~O_2^{-2}~  |\psi_{s\prime}^{x\prime}\rangle\\
& = & (1-\delta_{s, s\prime})  \left(21 \alpha_{-1}^x \alpha_{-1}^{x\prime} -3 \sqrt{30}(\alpha_2^x \alpha_{-4}^{x\prime} + \alpha_2^{x\prime} \alpha_{-4}^x)\right).
\end{eqnarray*}
Accordingly, the coupling of the transverse phonons to the CEF doublets can occur through both the quadrupolar operators $O_2^{\pm1}$ and $O_2^{\pm2}$ with oscillator strengths depending on the actual structure of the doublet states, that is, on the parameters $ \alpha_{M}^x$. Using numerical values published in the literature for the $\alpha_M$ coefficients \cite{Mirebeau07}, one observes that the strongest coupling comes from the $O_2^{\pm1}$ quadrupoles. Interestingly, the same numerical calculations performed for \hoti\, 
show that those quadrupolar matrix elements are essentially zero.
This is also the case for \dyti\, when generalizing the above analysis to Kramers ions.

In \TTO, an optical phonon is also suspected to exist at energy close to the first CEF excited doublet, which would belong to the irreducible representation  $T_{2u}$ of the factor group $m\bar{3}m$ (silent in both IR and Raman spectroscopy). Its subduction to the point group $D_{3d}$ is reduced into ${T_{2u}}_{\downarrow D_{3d}} = A_{1u} \oplus E_u$, meaning again that the associated triply degenerate phonon modes give rise to a singlet ($A_{1u}$), the longitudinal mode, and a doublet ($E_u$), the transverse modes. These transverse modes can couple to the first CEF excited doublet with similar oscillator strengths as calculated above resulting in the splitting of the CEF excited state. The longitudinal mode can also couple to the first CEF state through the quadrupolar operator $O_2^0$. No splitting will result but a shift in energy will occur. This is most likely the process that it activated below 200 K in \TTO.

\begin{table}
\begin{tabularx}{\linewidth}{p{0.7cm}p{0.7cm}p{0.7cm}p{0.7cm}p{0.7cm}p{0.7cm}p{0.7cm}p{4cm}*{1}{>{\arraybackslash}X}}
\hline
\hline
$\mathcal{D}_{3\mathrm{d}}$ &
$\bar{3}\mathrm{m}$ &
$\mathfrak{C}_1$ &
$\mathfrak{C}_2$ &
$\mathfrak{C}_3$ &
$\mathfrak{C}_{\bar{1}}$ &
$\mathfrak{C}_{\bar{2}}$ &
$\mathfrak{C}_{\bar{3}}$ &
$\mathrm{Basis~Vectors}~ - ~\mathrm{Invariants}$\\
\hline
$A_{1g}$ & $\Gamma_1^+$ & 1 & 1 & 1 & 1 & 1 & 1 & $Y_2^0$ \\
$A_{2g}$ & $\Gamma_2^+$ & 1 & -1 & 1 & 1 & -1 & 1 & $J_z$ \\
$E_{g}$  & $\Gamma_3^+$ & 2 & 0 & -1 & 2 & 0 & -1 & $(Y_2^{+m} \pm Y_2^{-m})_{(m=1,2)} ~ (J_x,~J_y)$ \\
$A_{1u}$ & $\Gamma_1^-$ & 1 & 1 & 1 & -1 & -1 & -1 & \\ 
$A_{2u}$ & $\Gamma_2^-$ & 1 & -1 & 1 & -1 & 1 & -1 & $Z$ \\
$E_{u}$ & $\Gamma_3^-$ & 2 & 0 & -1 & -2 & 0 & 1 & $(X,~Y)$\\
\hline
\hline
\end{tabularx}
\caption{Irreducible representations of the group $\mathcal{D}_{3\mathrm{d}}$, denoted $\Gamma_i^{\pm}$ and according to the Mulliken convention \cite{Mulliken55}, $\mathfrak{C}_1$ stands for the identity, $\mathfrak{C}_n$ for the class of the n-fold rotations, $\mathfrak{C}_{\bar{1}}$ for the space inversion and $\mathfrak{C}_{\bar{n}}$ for the class of the n-fold rotoinversion. The basis vectors (or the invariants) associated with each irreducible representation $\Gamma_i^\pm$ can be computed by means of the projectors $\mathcal{P}_i^\pm = \frac{\mathrm{dim}(\Gamma_i)}{\mathrm{Card(D_{3d})}}\sum_{\mathrm{g} \in \mathrm{D_{3d}}} \chi_i^\star(\mathrm{g})~\mathrm{g}$, where $\chi_i^\pm$ is the character of $\Gamma_i^\pm$, and when $\mathrm{dim}(\Gamma_i) > 1$ the exchangers $\mathcal{Q}_i^{nm} = \frac{\mathrm{dim}(\Gamma_i)}{\mathrm{Card(D_{3d})}}\sum_{\mathrm{g} \in \mathrm{D_{3d}}} {\Gamma_i^\pm}^{mn}(\mathrm{g^{-1}})~\mathrm{g}$ by applying these operators on trial objects (vector components, spherical functions).}
\label{D3d_IRREPS}
\end{table}

\subsection{Crystal electric field transitions\label{sec:trans}}

\begin{figure}
\centering
\includegraphics[width=0.5\linewidth]{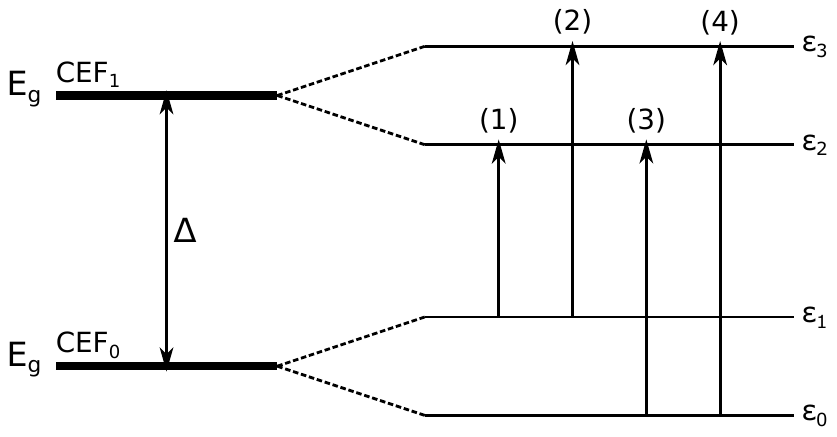}
\caption{Four possible transitions between ground, and first excited $E_g$ CEF doublets.}
\label{fig:Transitions}
\end{figure}

Considering a splitting of the ground and first excited CEF doublets, we can expect four possible transitions as seen in Fig.~\ref{fig:Transitions}. In the main article, we assign three of the observed peaks (P0, P1 and P2) within this transition scheme.  P0 corresponds to transition (1), P1 to transitions (2) and (3), and P2 is (4). P3 does not fit into the scheme and is therefore described by the $T_{2u}$ coupling mechanism.  However, this proposed scheme requires that the difference in  energy splitting of each doublet is equivalent ($\varepsilon_1-\varepsilon_0=\varepsilon_3-\varepsilon_2$). In such a case, transitions (2) and (3) will have the same energy difference and therefore appear as one peak. If we assume different energy splittings ($\varepsilon_1-\varepsilon_0\neq\varepsilon_3-\varepsilon_2$), then it should be possible to observe the four distinct transitions. We may then expect that the four peaks observed in the main article correspond to the four transitions here. While the different temperature profiles of peaks P0, P1 and P2 compared to P3 is a strong indication that P3 does not fit into this scheme, here we show arithmetically that such a four peak scheme is unlikely given their observed energies.

First we define $\varepsilon_0 = 0$. Then the largest transition (4) is $\varepsilon_0 \rightarrow \varepsilon_3$.  If we take the highest energy peak P3 = 0.675\,THz  as this transition, then $\varepsilon_3$ = 0.675\,THz. Now, depending on whether  $\varepsilon_1-\varepsilon_0>\varepsilon_3-\varepsilon_2$ or $\varepsilon_1-\varepsilon_0<\varepsilon_3-\varepsilon_2$, the next largest transition will be either (3) $\varepsilon_0 \rightarrow \varepsilon_2$ or (2) $\varepsilon_1 \rightarrow \varepsilon_3$. Assuming the first case, P2 = 0.50\,THz corresponds to $\varepsilon_0 \rightarrow \varepsilon_2$ setting $\varepsilon_2$ = 0.50\,THz. So $\varepsilon_3-\varepsilon_2$ = 0.175\,THz. Now, (2) is the next largest transition and must correspond to P1 = 0.414\,THz. So $\varepsilon_3-\varepsilon_1$ = 0.414\,THz. Since we have shown $\varepsilon_3$ = 0.675\,THz, then $\varepsilon_1$ = (0.675 - 0.414)\,THz = 0.261\,THz. Therefore, transition (1) which is $\varepsilon_1 \rightarrow \varepsilon_2$ must be (0.50 - 0.261)\,THz = 0.239\,THz which is not consistent with our measurement of P0 = 0.33\,THz. The same outcome occurs when considering the case $\varepsilon_1-\varepsilon_0<\varepsilon_3-\varepsilon_2$. Therefore, a scheme that attempts to conform the four observed peaks into the four transitions in Fig.~\ref{fig:Transitions} cannot reconcile with the energies in our spectra.
Furthermore, such a scheme is also inconsistent with previous inelastic neutron scattering studies.

\subsection{Detailed experimental methods\label{sec:methods}}
\subsubsection{Sample details}
\begin{figure}
\includegraphics{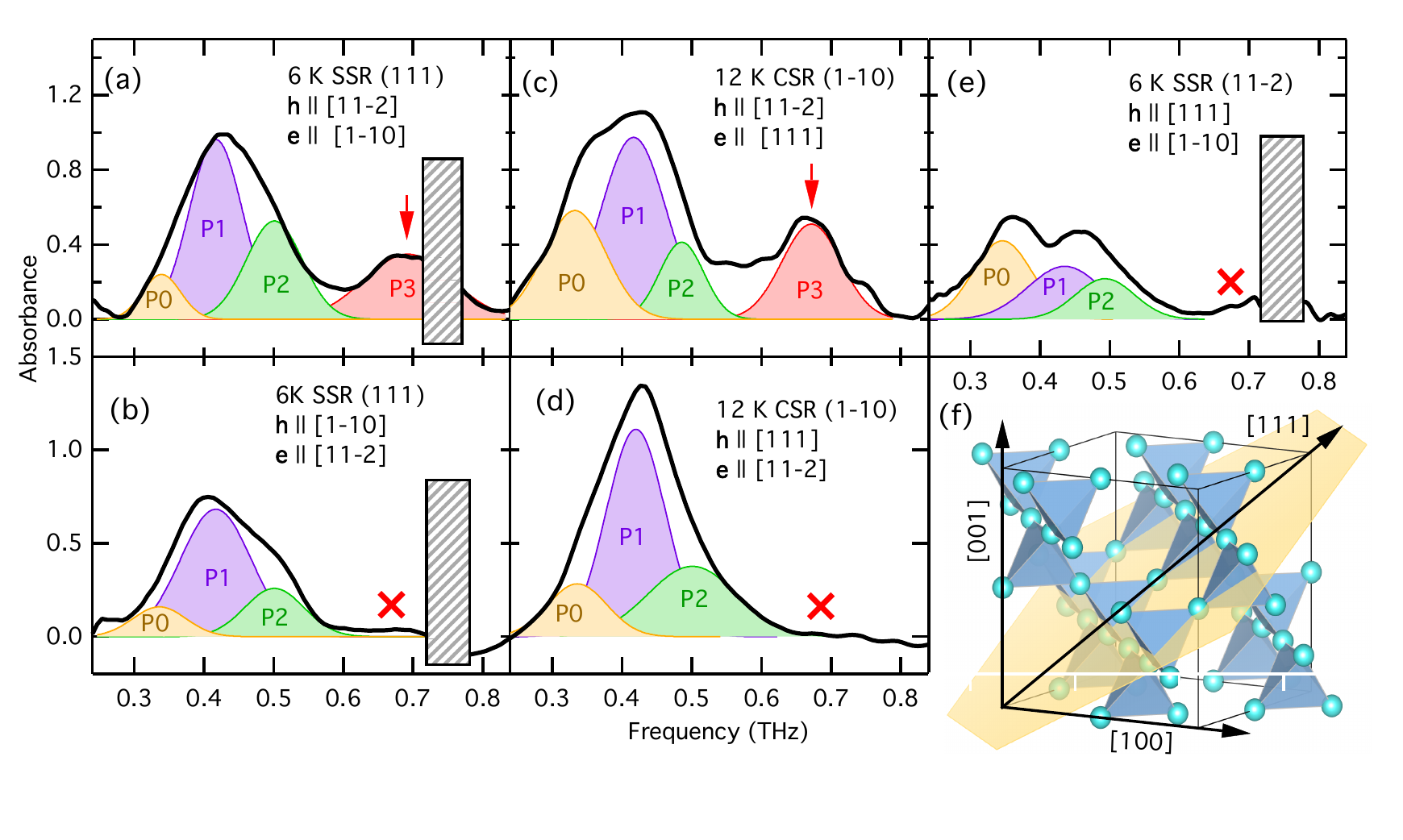}
 \caption{(a)-(e) Polarization dependent absorbance spectra recorded at 6\,K or 12\,K with SSR or CSR configurations. The hatched rectangles mask the beamsplitter artifact at 0.75\,THz in the SSR spectra. The peak fits for P0-P3 are shown in each spectra identifying that the multi peak structure is indeed reproducible. Red arrows and crosses indicate orientations where the P3 excitation is present and absent respectively (f) The (11$\bar{2}$) plane within the pyrochlore lattice. Aligning $\boldmath{h}$ in this plane produces the strongest absorption from the P3 excitation.}
 \label{fig:SlctnRls}
 \end{figure}

The single crystal samples used in the experiments were cut from a larger crystal used in a previous neutron study \cite{Guitteny2013}.
They were then prepared into thin wafers.
Each sample was selected with a specific crystallographic face confirmed and orientated by Laue reflected X rays.
Three mutually orthogonal planes (111), (1$\bar{1}$0) and (11$\bar{2}$) were chosen to test the electro/magneto activity of the excited modes.
With polarized radiation incident on the (111) plane, the magnetic ($\textbf{h}$) and electric ($\textbf{e}$) field components can probe either the (1$\bar{1}$0) or (11$\bar{2}$) planes.
Investigating the (1$\bar{1}$0) cut sample allows $\textbf{h}$ and $\textbf{e}$ field probing of the (111) and (11$\bar{2}$) planes.
In principle, the $\textbf{h}$ field can therefore be aligned with the (11$\bar{2}$) plane in both cases while the $\textbf{e}$ field probes different planes.
In this way, it is possible to isolate the effects of $\textbf{h}$ or $\textbf{e}$ fields on specific planes in the crystal.
An example of this is shown in Fig~\ref{fig:SlctnRls} where $\textbf{h}$ is isolated along the [11$\bar{2}$] direction by comparing polarized spectra of the three orthogonally cut samples.
Here, it is clear that the absorption from the P3 mode is strongest when $\boldmath{h}$ lies in the (11$\bar{2}$) plane.
A final sample cut normal to the low index (100) plane was also selected to compare with any observed spectroscopic anisotropies.
The (111), (1$\bar{1}$0) and (11$\bar{2}$) cut samples had a thickness of 0.2\,mm, 0.3\,mm and 0.3\,mm respectively.
The (100) sample had a thickness of 1\,mm.

\subsubsection{Terahertz synchrotron setup}
The THz spectroscopy experiments were performed at the SOLEIL synchrotron in France on the AILES Far-Infrared beamline.
Spectra were obtained in the 0.15--1.0\,THz (0.62--3.8\,meV) energy range with a Br\"{u}ker IFS125 spectrometer and 1.6\,K bolometer using both standard (SSR) and coherent synchrotron radiation (CSR) techniques.
In the studied energy range, the THz wave is linearly polarized at more than 95\%. 
For the SSR measurements,
 a 125\,$\mu$m mylar beam splitter was used to access the low energies required for the experiment.
Multiple reflections in the beam splitter produce a spectral minimum at0.75\,THz, therefore, for SSR measurements, we omit data at this frequency due to a high noise level.
In this configuration, samples were mounted at fixed orientations with respect to the THz polarization on the cold head of an adapted Cryomech PT 405 pulse tube cryostat.
The temperature range accessible was 6--300\,K.
In the CSR configuration, the  SOLEIL accelerator was operating in the low-$\alpha$ coherent mode, with a $\alpha/\alpha_0$ parameter of 1/25 \cite{CSR_1,CSR_2}.
Under this operation,
a much higher flux of low-energy photons is produced.
This facilitated the use of a 50\,$\mu$m mylar beam splitter that removes the spectral minima at  0.75\,THz while maintaining strong signal at low energies.
In this configuration, samples were mounted on an Attocube ANR240 rotary stage attached to the cold head of the cryostat.
The thermalization of the rotary stage limited the base temperature to 12\,K.
Using the rotary stage, the sample could be rotated through 100\degree with respect to the THz polarization with a precision of 0.1\degree.
During analysis the data were filtered using a Fourier filtering procedure that removes the 2\,\wvnmbr fringe noise due to multiple reflections in the setup \cite{FTIR_filtr}.

\subsubsection{Far-infrared setup}

Far-infrared measurements of the optical phonon spectra were performed in reflection geometry at near normal incidence.
The measurements were performed on a Br\"{u}ker Vertex 70v spectrometer with a mercury discharge-lamp source and 4.2\,K bolometer detector.
A far-infrared silicon  beam-splitter was used, with the measurements covering the 50--900\,\wvnmbr (1.5--27\,THz, 6--112\,meV) spectral range.
The sample was mounted on the cold head of an in-house--adapted Janis flow cryostat reaching temperatures from 6 to 300\,K



\subsection{Far-infrared phonon spectra\label{sec:FIR}}

  \begin{figure}
 \centering
 \includegraphics{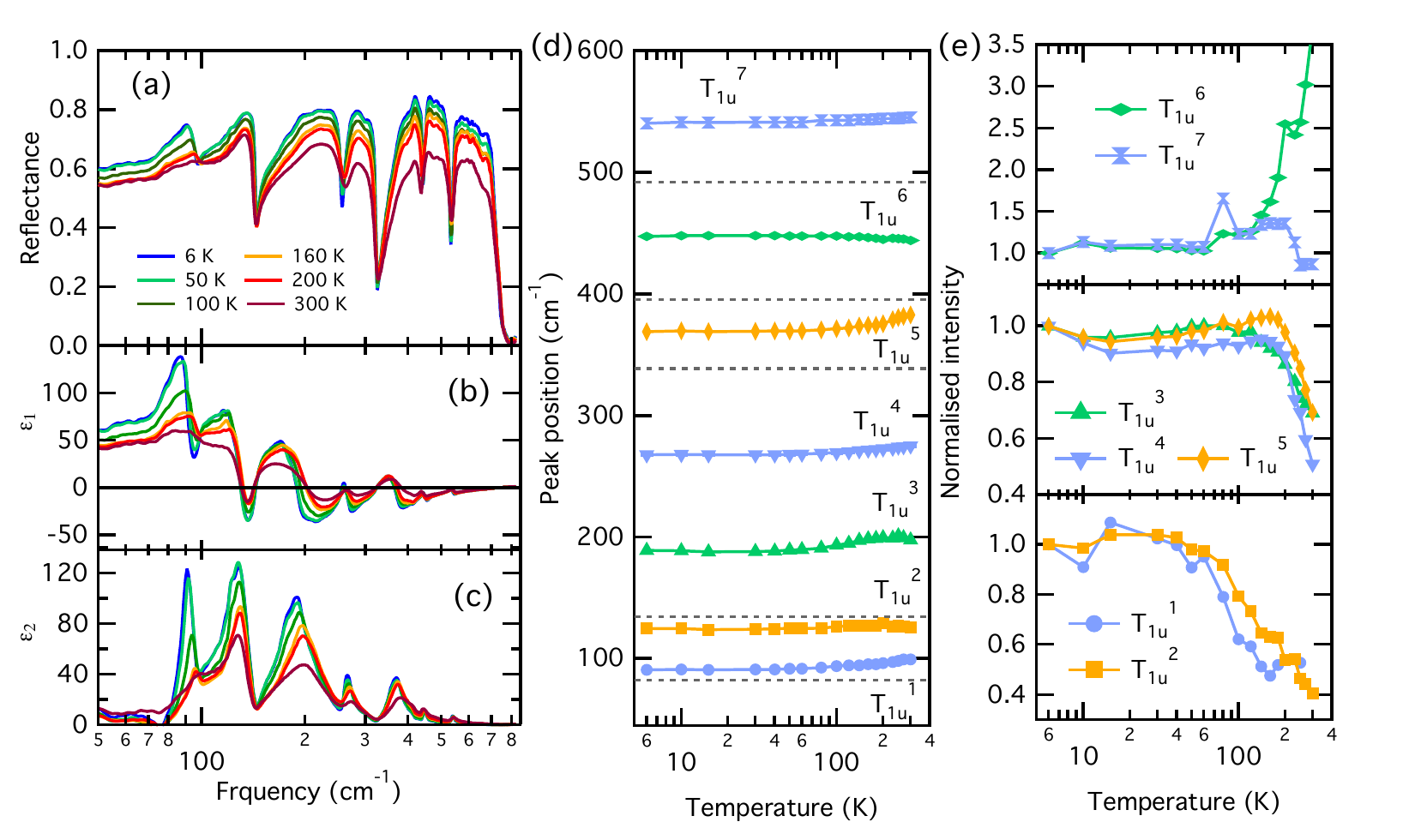}
 \caption{(a) Far-infrared reflectivity of (100) cut sample from 6--300\,K. (b) Real and (c) imaginary components of dielectric function determined by Kramers-Kronig analysis. Peaks in $\epsilon_2$ represent the absorption bands of phonons. (d) Temperature dependence of fitted phonon frequencies. Dashed gray lines indicate energies of known crystal field levels. (e) Temperature dependence of the normalized peak intensities.}
 \label{fig:Phonon}
 \end{figure}
To complement the terahertz spectroscopy provided in the article, we have also performed far-infrared reflectance spectra of the (100) cut sample.
Here we aim to analyse the infrared active optical phonon spectra and their temperature dependence to gain insight into the lattice dynamics.
The reflectance spectra at selected temperatures from 6--300\,K are shown in Fig.~\ref{fig:Phonon} (a).
Here, a number of Reststrahlen bands are present due to phonon absorption. The $A_2B_2O_6O'$ stoichiometry of \TTO tells us that the
 $A^{3+}$ and $B^{4+}$ ions occupy the 16d and 16c Wyckoff positions while the $O$ and $O'$ ions are at the 48$f$ and 8$b$ positions respectively.
Factor group analysis of the $m3\bar{m}$ point group table revels the allowed normal modes at the zone center, which are given by the following representations \cite{BilbaoSAM},
 \begin{equation*}
 \Gamma = A_{g1} \oplus 3A_{2u} \oplus 3E_u \oplus E_g \oplus 4T_{2u} \oplus 4T_{2g} \oplus 8T_{1u} \oplus 2T_{1g}\quad.
 \end{equation*}
 Of these modes, only the 8 $T_{1u}$ modes are infrared active with one being an acoustic mode.
 We therefore expect 7 phonon modes to appear in our reflectivity data.
 Because of the numerous crystal field levels known to be mixed within the phonon spectrum and 
 our symmetry analysis allowing
 coupling between them, the precise form of the dielectric response is non-trivial.
 Rather than applying the dispersive method of fitting the reflectivity to a simplified dielectric model (as in the analysis of Bi \textit{et al.} for Dy$_2$Ti$_2$O$_7$ \cite{Bi}), here we analyze the dielectric response by Kramers-Kronig analysis.
 The real and imaginary components of the dielectric response, determined by a Kramers-Kronig transform of the reflectivity, are shown in Figs.~\ref{fig:Phonon} (b) and (c).
Peaks in the imaginary response represent absorptions due to phonon excitations.
 Seven distinct peaks are observed corresponding to the 7 predicted $T_{1u}$ optical modes.
No soft modes are present which may have indicated a structural instability or phase transition.
 Notably, the lowest energy phonons ($T^1_{1u}$, and  $T^2_{1u}$) in close proximity to the 2nd and 3rd crystal field levels, feature abnormal peak profiles and temperature dependence.
 Nevertheless, approximate phonon energies and intensities can be extracted by Lorentzian profile fits to the curves.
 The temperature dependence of the peak positions are shown in Fig.~\ref{fig:Phonon} (d), with dashed gray lines representing the neighboring crystal field levels.
 The normalized phonon intensities, determined by the area of the fitted peaks, are shown in Fig.~\ref{fig:Phonon} (e).
 While the 5 higher energy phonons follow a similar temperature dependence in their intensity profile, strengthening or weakening as the lattice contracts between room temperature and 200\,K, the two lower energy modes are distinct. These two modes, which are suspected of interacting with the crystal field levels, feature rapid strengthening between 200\,K and 80\,K.
 Notably this is similar to the temperature dependence of P3 from Fig.~4 (c). 
 A summary of the results at base temperature is presented in Table~\ref{Tbl:Peaks}.

 \begin{table}[h]

 \def\arraystretch{1.5}
 \caption{Table of transverse optical phonon energies determined by spectra at base temperature (6\,K) compared with previously reported results.
 }
 \label{Tbl:Peaks}
 \begin{tabular} {r|ccc}
 \hline \hline
  Symmetry& This work & Ruminy \textit{et al.} \cite{Ruminy1} & Bi \textit{et al.} (\dyti) \cite{Bi} \\
\hline
Phonon&\wvnmbr&\wvnmbr&\wvnmbr\\
$T_{1u}$  & 90.9 	& 62.1 	& 85	\\
          & 124.9 	& 117.9	& 126 	\\
		  & 189.1 	& 176.2	& 197	\\
		  & 267.8 	& 230.2	& 260 	\\
		  & 368.9 	& 383.1	& 370 	\\
		  & 447.4 	& 452.7	& 454 	\\
		  & 540.3 	& 536.8	& 545 	\\ \hline \hline
\end{tabular}
\end{table}

\bibliographystyle{apsrev4}

\end{document}